\documentclass[12pt]{iopart}
\pdfoutput=1
\usepackage{graphicx}
\usepackage{amsfonts}
\usepackage{amssymb}

\begin{document}

\title{Dissipative preparation of generalised Bell states}
\author{R Sweke, I Sinayskiy and F Petruccione}
\address{Quantum Research Group, School of Chemistry and Physics, and National Institute for Theoretical Physics, University of KwaZulu-Natal, Durban, 4001, South Africa.}
\eads{\mailto{rsweke@gmail.com}, \mailto{sinayskiy@ukzn.ac.za}, \mailto{petruccione@ukzn.ac.za}}

\begin{abstract}
A scheme is presented for the dissipative preparation of generalised Bell states of two-qubits, within the context of cavity QED. In the suggested protocol the dissipative processes of spontaneous emission and cavity loss are no longer undesirable, but essential to the required dynamics. Extremely long lived target states are achieved, with fidelities of near unity, utilising cooperativities corresponding to currently available optical cavities. Furthermore, the suggested protocol exhibits excellent scaling of relevant characteristics, with respect to cooperativity, such that improved results may be obtained as the development of experimental capabilities continues.
\end{abstract}

\pacs{03.67.Bg, 42.50.Pq, 42.50.Dv}

\maketitle

\section{Introduction}\label{Intro}

A path towards the experimental realization of a quantum computer has become one of the main focus areas of current quantum information research \cite{compreal1}. Since Shor's ground breaking  algorithm for factorisation \cite{ALG1} many quantum algorithms have been designed and studied \cite{ALG2,ALG3}. However, in order for the implementation of these algorithms to become a reality it is essential to be capable of creating and manipulating large scale entanglement between effective physical qubits. One of the primary obstacles in this regard is the interaction of a system with its environment, resulting in dissipation and decoherence \cite{FRAN}. Classical computer science has provided an effective strategy for combating these destructive effects, on unitary implementations of quantum algorithms, in the form of error-correcting codes \cite{QEC1} - \cite{QEC3}. This approach, based on treating the system-environment interaction as a negative influence, the effect of which needs to be minimized, has been further refined through the introduction of fault tolerant computation \cite{FTQC1}-\cite{FTQC3}. Recently, the culmination of this approach has been achieved in the thresh-hold theorems \cite{THR1,THR2} which now provide an intelligent measure of our progress towards a large scale quantum computer.

However, a paradigm shift in the approach towards the physical realisation of a quantum computer has recently been introduced. This shift has arisen as a result of the theoretical prediction that dissipation can in fact be utilized for the creation of complex entangled states \cite{DPREPprin1}-\cite{DPREPprin4} and to perform universal quantum computation \cite{DCOMP1}-\cite{DCOMP3}. This fundamental shift in approach is based on the assumption that the system environment coupling can be manipulated such that the system is driven towards a steady state which is the solution to a computational task, or a desired entangled state \cite{DPREPprin1}. Within this approach dissipation is no longer a negative effect, but crucial to the required dynamics. 

Since the introduction of dissipative state preparation and dissipative quantum computing \cite{DPREPprin1}-\cite{DPREPprin4}, a large amount of effort has been put into the physical realisation of these ideas, within a broad range of physical systems. In particular, Cavity QED setups have received a lot of attention \cite{DPREPQED1}-\cite{DPREPQED7}, while suggestions have also been made for implementation within atomic ensembles \cite{DPREPAE}, trapped ions \cite{DPREPTI} and NV centres in diamond \cite{DPREPNV}. Importantly, two recent experimental realisations of dissipative state preparation, within macroscopic ensembles \cite{DPREPREAL2} and trapped ions \cite{DPREPREAL1}, have validated the importance of this approach and provided an impetus for further research.

Furthermore, since the ground breaking realisation of a two-bit quantum logic gate in 1995 \cite{EQED1,EQED2}, the experimental progress within cavity QED has been vast \cite{qedreview3}, and the obstacles presented by dissipation and decoherence well understood. This progress, coupled with the mass of theoretical progress on dissipative state preparation within cavity QED \cite{DPREPQED1}-\cite{DPREPQED7}, motivates the continued study of dissipative state preparation and quantum computing within the context of cavity QED. 

In this paper we suggest a physically realisable scheme for the dissipative preparation of generalised Bell states \cite{EPR}, 

\begin{eqnarray}\label{arbtargstates}
|\psi^{+}\rangle \equiv \mathrm{cos}(\theta) |1 0\rangle + \mathrm{sin} (\theta) |0 1\rangle \label{arbtarg1} \\
|\psi^{-}\rangle \equiv \mathrm{cos}(\theta) |0 1\rangle - \mathrm{sin} (\theta) |1 0\rangle \label{arbtarg2}
\end{eqnarray}
for arbitrary values of $\theta$. For the purposes of this paper we utilise ensembles of $\Lambda$ atoms within a single optical cavity, as per Figure \ref{arbstatesetup}. Insight into dissipative mechanisms for the preparation of the generalised Bell states (\ref{arbtarg1}) and (\ref{arbtarg2}) is an important step forward for practical dissipative state preparation, as well as for the development of methods of control for complex open quantum systems.

As in previous suggestions for the dissipative preparation of the maximally entangled state of two qubits \cite{DPREPQED1,DPREPQED2, DPREPQED6} and a W state of three qubits \cite{DPREPQED5}, we implement adiabatic elimination, via the effective operator formalism \cite{effop}, in order to obtain a reduced master equation. The structure of this reduced master equation allows for the identification of effective decay channels, allowing cavity parameters to be engineered such that the target state is prepared efficiently and reliably.  Numerical results show that extremely long lived target states can be achieved, utilising cooperativities corresponding to currently available optical cavities. Furthermore, the suggested scheme exhibits excellent scaling characteristics with respect to cooperativity, such that with the development of higher cooperativity cavities \cite{qedreview3}, even better results will be physically achievable.

\section{Physical set-up and effective operators}

\begin{figure}
\begin{center}
\includegraphics[width=0.7\linewidth]{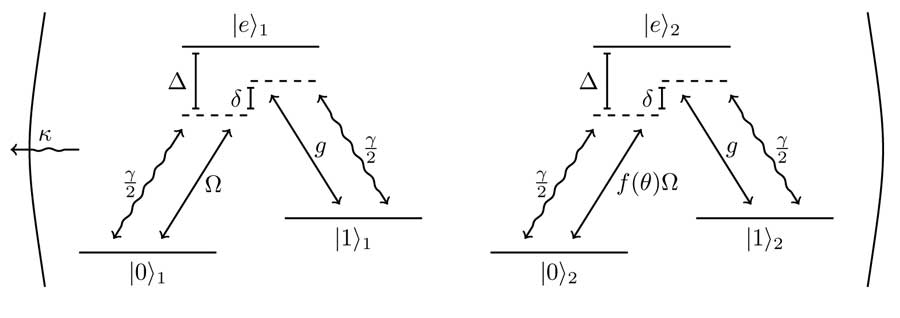} 
\caption{Cavity QED setup for the dissipative preparation of an arbitrary two-qubit state between two $\Lambda$ atoms in a single-mode optical cavity. The $i$'th $\Lambda$ atom consists of two ground states $(|0\rangle_{i}, |1\rangle_{i})$ and an excited state $|e\rangle_{i}$. Each atom experiences individual coherent driving, coupling the levels $|0\rangle$ and $|e\rangle$, as well as an atom-cavity interaction coupling the levels $|1\rangle$ and $|e\rangle$. The system interacts with the environment via spontaneous emission and cavity loss.}\label{arbstatesetup}
\end{center}
\end{figure}

We consider a Cavity QED setup of two distant $\Lambda$ atoms in a single-mode optical cavity, as per Figure \ref{arbstatesetup}. Each $\Lambda$ atom consists of two ground states, $|0\rangle$ and $|1\rangle$, and an excited state $|e\rangle$, coupled to the cavity mode. The Hamiltonian for the system is given  by

\begin{equation} \label{hamform}
\hat{H} = \hat{H}_{g} + \hat{H}_{e} + \hat{W}_{+} + \hat{W}_{-},
\end{equation}
where $\hat{H}_{e}$ is the Hamiltonian for the excited subspace, $\hat{H}_{g}$ the Hamiltonian for the ground subspace, $\hat{W}_{+}$ the perturbative excitation from the ground space to the excited space and $\hat{W}_{-}$ the perturbative de-excitation. As required by adiabatic elimination \cite{adelim1}, which we wish to implement, we restrict ourselves to the single-excitation subspace and find that in the appropriate rotating frame, and the conventional computational basis, the Hamiltonian is time independent with individual terms given by

\begin{eqnarray}
\hat{H}_{\mathrm{e}} = \Delta(|e\rangle_{1}\langle e| + |e\rangle_{2}\langle e|) + \delta \hat{a}^{\dagger}\hat{a} + \hat{H}_{\mathrm{ac}}, \label{one}\\
\hat{H}_{\mathrm{ac}} = g\hat{a}(|e\rangle_{1}\langle 1| + |e\rangle_{2}\langle 1|) + \mathrm{H.c.}, \\
\hat{W}_{+} = \frac{\Omega}{2} (|e\rangle_{1}\langle 0| + f (\theta)|e\rangle_{2}\langle 0|), \\
\hat{W}_{-} = \hat{W}_{+}^{\dagger},\\
\hat{H}_{\mathrm{g}} = 0, \label{five}
\end{eqnarray}
where $f(\theta)$ is as of yet undefined and $\hbar = 1$. As one wishes to prepare an arbitrary state, it is necessary to obtain system dynamics not restricted to a single irreducible subspace of the total Hilbert space and hence one requires individual addressing of atoms within the cavity, as opposed to uniform global addressing. In order to achieve this the perturbative excitation $\hat{W}_{+}$ is driven by two coherent lasers, the first laser driving the first atom with Rabi frequency of $\Omega$ and a detuning of $\Delta$, and the second laser driving the second atom with a Rabi Frequency of $f(\theta)\Omega$ but with the same detuning. The atom-cavity interaction, described by $\hat{H}_{\mathrm{ac}}$, couples the levels $|e \rangle$ and $|1\rangle$ with a strength of $g$ and a uniform phase over both atoms.

We assume Markovian interaction with the environment, an excellent assumption within quantum optics \cite{carQO}, and in addition assume that the system is at zero absolute temperature (where it is important to note that this assumption is not an obstacle to experimental implementation \cite{qedreview3}). As a result, the system evolves according to a master equation of the form
  
\begin{equation}\label{masterequation}
 \dot{\rho} = \mathcal{L}\rho = -i[\hat{H},\rho]\nonumber  + \sum_{k}\left(\hat{L}_{k}\rho \hat{L}^{\dagger}_{k} - \frac{1}{2}\hat{L}_{k}\hat{L}^{\dagger}_{k}\rho - \frac{1}{2}
\rho \hat{L}^{\dagger}_{k}\hat{L}_{k}\right),
\end{equation}
where $\hat{L}_{k}$ are Lindblad operators describing the interaction of the system with the environment via cavity loss and spontaneous emission. In the system under consideration, as illustrated in Figure \ref{arbstatesetup}, there are five distinct decay processes. The first decay process, that of cavity loss, is described by Lindblad operator $\hat{L}_{1}$, given by

\begin{equation}
\hat{L}_{1} = \sqrt{\kappa}\hat{a}.
\end{equation}
Lindblad operators $\hat{L}_{2}$ and $\hat{L}_{3}$, given by

\begin{eqnarray}
\hat{L}_{2} = \hat{L}_{[\gamma,(1,0)]} = \sqrt{\frac{\gamma}{2}}|0\rangle_{1}\langle e|, \\
\hat{L}_{3} = \hat{L}_{[\gamma,(2,0)]} = \sqrt{\frac{\gamma}{2}}|0\rangle_{2}\langle e|,
\end{eqnarray}
then describe spontaneous emission of the first and second atom respectively into the ground state $|0\rangle$ (where the notation $\hat{L}_{[\gamma,(i,j)]}$ is used to describe spontaneous emission of the $i$'th atom from state $|e\rangle_{i}$ into state $|j\rangle_{i}$, with $j \in \{0,1\}$). Finally, Lindblad operators $\hat{L}_{4}$ and $\hat{L}_{5}$, given by

\begin{eqnarray}
\hat{L}_{4} = \hat{L}_{[\gamma,(1,1)]} = \sqrt{\frac{\gamma}{2}}|1\rangle_{1}\langle e|, \\
\hat{L}_{5} = \hat{L}_{[\gamma,(2,1)]} = \sqrt{\frac{\gamma}{2}}|1\rangle_{2}\langle e|,
\end{eqnarray}
describe spontaneous emission of the first and second atom respectively into the ground state $|1\rangle$. 
For simplicity, decay rates into states $|0\rangle$ and $|1\rangle$ have been set equal, however this is not a necessary requirement \cite{DPREPQED1, DPREPQED2}. In order to simplify our description of the system and facilitate the identification of effective decay channels, it is extremely useful to adiabatically eliminate both atomic and cavity excitations. We apply adiabatic elimination (whose requirements are discussed shortly), via the effective operator formalism \cite{effop}, in order to obtain a reduced master equation of the form

\begin{equation}\label{effmastereq}
 \fl\dot{\rho}_{g} = -i[\hat{H}_{\mathrm{eff}},\rho_{g}]  + \sum_{k}\Big(\hat{L}^{k}_{\mathrm{eff}}\rho_{g} (\hat{L}^{k}_{\mathrm{eff}})^{\dagger}  
 -  \frac{1}{2}\hat{L}^{k}_{\mathrm{eff}}(\hat{L}^{k}_{\mathrm{eff}})^{\dagger}\rho_{g} - \frac{1}{2}
\rho_{g} (\hat{L}^{k}_{\mathrm{eff}})^{\dagger}\hat{L}^{k}_{\mathrm{eff}} \Big),
\end{equation}
where $\rho_{g}$ is the density matrix for the \emph{ground subspace}. The effective operators are given by

\begin{eqnarray}
\hat{H}_{\mathrm{eff}} \equiv- \frac{1}{2}\hat{W}_{-}\Big( \hat{H}_{NH}^{-1} + (\hat{H}_{NH}^{-1})^{\dagger}\Big)\hat{W}_{+} + \hat{H}_{g},  \label{heffdef}\\
\hat{L}^{k}_{\mathrm{eff}} \equiv \hat{L}_{k} \hat{H}_{NH}^{-1} \hat{W}_{+}, \label{leffdef}
\end{eqnarray}
with $H_{NH}$, the non-Hermitian Hamiltonian combining real detunings of excited levels with imaginary terms corresponding to irreversible decay, given by,

\begin{equation}\label{HNHdef}
\hat{H}_{NH} \equiv \hat{H}_{e} - \frac{i}{2}\sum_{k}\hat{L}_{k}^{\dagger}\hat{L}_{k}.
\end{equation}
In order to apply this formalism we will work within the high cooperativity regime $g^2 \gtrsim \kappa\gamma$ and in addition, in order to apply adiabatic elimination (and motivate a restriction to the single-excitation subspace), it is required that we restrict ourselves to the regime of weak driving $(\Omega,f(\theta)\Omega) \ll (g,\kappa,\gamma)$ and simultaneously ensure that the excited energy levels are largely detuned from the ground levels, i.e., that $\Delta$ (the detuning of the coherent interaction between $|0\rangle$ and $|e\rangle$) and $\Delta - \delta$ (the detuning of the atom-cavity interaction between $|1\rangle$ and $|e\rangle$) are both large, implying $(\Delta,\Delta - \delta) \sim g$. 

The relevant subspace of the total Hilbert space for this physical situation is

\begin{equation}\label{hilbarb}
\mathcal{H} = \mathbb{C}^3\otimes\mathbb{C}^3\otimes\mathbb{C}^2,
\end{equation}
the correct subspace for two $\Lambda$ atoms within a single-mode optical cavity restricted to the single-excitation subspace. As we wish to prepare states of the form given in (\ref{arbtargstates}), it is convenient to move into the basis

\begin{equation}\label{basis}
B = \{G,A,C\},
\end{equation}
where $G$ is the basis for the ground state subspace of (\ref{hilbarb}), given by

\begin{equation}
G = \Big\{|00\rangle, |\psi^{+}\rangle, |\psi^{-}\rangle, |11\rangle \Big\},
\end{equation}
and for all states $|\alpha\rangle$ in $G$ we have naturally defined

\begin{equation}
|\alpha\rangle \equiv |\alpha\rangle\otimes|0\rangle,
\end{equation}
with the states in the tensor product describing the atomic state of both atoms and the cavity state respectively. $A$ is the basis for the subspace of states with a single atomic-excitation, given by

\begin{equation}\label{additional}
A = \Big\{|\psi_{0}^{+}\rangle, |\psi_{0}^{-}\rangle, |\psi_{1}^{+}\rangle, |\psi_{1}^{-}\rangle \Big\},
\end{equation}
where we have defined

\begin{eqnarray}
|\psi_{j}^{+}\rangle \equiv \mathrm{cos}(\theta)\big(|e\rangle\otimes|j\rangle\otimes|0\rangle\big) + \mathrm{sin}(\theta)\big(|j \rangle \otimes|e\rangle\otimes|0\rangle\big), \\
|\psi_{j}^{-}\rangle \equiv \mathrm{cos}(\theta)\big(|j\rangle\otimes| e\rangle\otimes|0\rangle\big) - \mathrm{sin}(\theta)\big(|e  \rangle\otimes|j\rangle\otimes|0\rangle\big),
\end{eqnarray}
with $j \in \{0,1\}$, following (\ref{additional}), and in correspondence with (\ref{hilbarb}) the triple tensor products describe the state of the first atom, second atom and cavity mode respectively.  Finally $C$ is the basis for the subspace containing states with a single cavity excitation, given by

\begin{equation}
C = \Big\{|00_{c}\rangle, |\psi_{c}^{+}\rangle, |\psi^{-}_{c}\rangle, |11_{c}\rangle \Big\},
\end{equation}
where we have utilized the natural definition 

\begin{equation}
|\alpha_{c}\rangle \equiv |\alpha\rangle\otimes|1\rangle,
\end{equation}
for all states $|\alpha\rangle$ in $G$. After a transformation into the basis $B$, the individual Hamiltonian terms and Lindblad operators are as per (\ref{transH})-(\ref{transL5}), given in Appendix A.

\begin{figure}
\begin{center}
\includegraphics[width=0.7\linewidth]{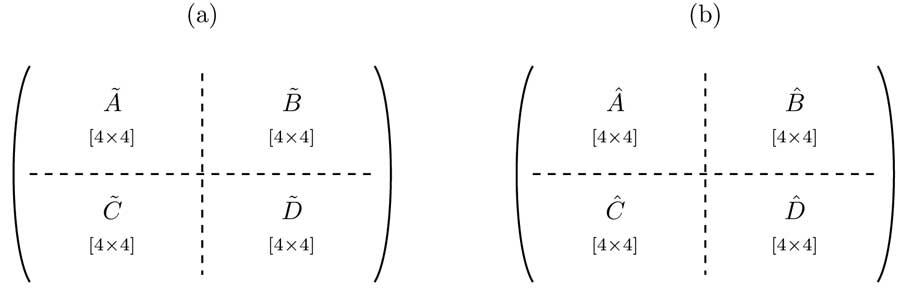} 
\caption{Partitioned matrix form of (a) $\hat{H}_{NH}$ and (b) $\hat{H}_{NH}^{-1}$ , corresponding to the physical set-up described by (\ref{one})-(\ref{five}). }\label{matarb}
\end{center}
\end{figure}

In this basis it is possible to utilise the effective operator formalism \cite{effop} to obtain effective operators between ground states, the form of which allows one to identify relevant decay channels and hence engineer cavity parameters such that the desired target state is prepared efficiently and reliably. Utilising (\ref{transH})-(\ref{transL5}) and (\ref{HNHdef}) one finds that the non-Hermitian matrix, $\hat{H}_{NH}$, can be written as a partitioned matrix, as per Figure \ref{matarb}. In this representation $\tilde{A}$ is the block pertaining to interactions within the single-cavity excitation subspace, $\tilde{D}$ is the block pertaining to interactions within the single atomic-excitation subspace and $\tilde{B}$,$\tilde{C}$ are blocks describing interactions between the two single-excitation subspaces. However, we are particularly interested in the inverse of $\hat{H}_{NH}$, the matrix of propagators between excited states, which via the Banachiewicz inversion theorem \cite{matrixinv} can also be written as a partitioned matrix, as in Figure \ref{matarb}, with individual blocks given by,

\begin{eqnarray}
\hat{D} = \big(\tilde{D} - \tilde{C}\tilde{A}^{-1}\tilde{B}\big)^{-1}, \label{invert1}  \\
\hat{A} = \tilde{A}^{-1} + \tilde{A}^{-1}\tilde{B}\big(\tilde{D} - \tilde{C}\tilde{A}^{-1}\tilde{B}\big)^{-1}\tilde{C}\tilde{A}^{-1}, \\
\hat{B} = -\tilde{A}^{-1}\tilde{B}\big(\tilde{D} - \tilde{C}\tilde{A}^{-1}\tilde{B}\big)^{-1}, \\
\hat{C} = \hat{B}^{T} = -\big(\tilde{D} - \tilde{C}\tilde{A}^{-1}\tilde{B}\big)^{-1}\tilde{C}\tilde{A}^{-1}.\label{invert2}
\end{eqnarray}
Utilising (\ref{invert1})-(\ref{invert2}) and defining

\begin{equation}
d_{n} = \bigg(\Delta - i\frac{\gamma}{2}\bigg)\bigg(\delta - i\frac{\kappa}{2}\bigg) - ng^2,
\end{equation}
and

\begin{equation}
\tilde{d}_{n} = 4ng^2 + (\gamma + 2i\Delta)(\kappa + 2i\delta).
\end{equation}
such that

\begin{equation}
d_{n}^{-1} = \frac{-4}{\tilde{d}_{n}},
\end{equation}
the blocks of $\hat{H}_{NH}^{-1}$ can be written as

\begin{eqnarray}
\fl\hat{A} = \bigg( \frac{-4\delta + 2i\kappa}{\tilde{d}_{1}}\bigg)|\psi_{0}^{\pm}\rangle\langle\psi_{0}^{\pm}|+ \bigg( \frac{g^2\mathrm{cos}(2\theta)}{d_{2}(\Delta - i\gamma/2)}\bigg)(|\psi_{1}^{-}\rangle\langle\psi_{1}^{+}| + |\psi_{1}^{+}\rangle\langle\psi_{1}^{-}|)\nonumber\\ 
 + \bigg( \frac{d_{2} + g^2(1 - \mathrm{sin}(2\theta))}{d_{2}(\Delta - i\gamma/2)}\bigg)|\psi_{1}^{-}\rangle\langle\psi_{1}^{-}| \nonumber\\
 + \bigg( \frac{d_{2} + g^2(1 + \mathrm{sin}(2\theta))}{d_{2}(\Delta - i\gamma/2)}\bigg)|\psi_{1}^{+}\rangle\langle\psi_{1}^{+}|,
\end{eqnarray}

\begin{eqnarray}
\fl\hat{B} = -g \bigg[\frac{1}{d_{1}}\Big(|\psi_{0}^{+}\rangle\langle\psi_{c}^{+}| + |\psi_{0}^{-}\rangle\langle\psi_{c}^{-}|\Big) 
 +\frac{1}{d_{2}}\Big((\mathrm{cos}(\theta) + \mathrm{sin}(\theta))|\psi_{1}^{+}\rangle\langle 11_{c}|  \nonumber\\ + (\mathrm{cos}(\theta) - \mathrm{sin}(\theta))|\psi_{1}^{-}\rangle\langle 11_{c}|\Big)  \bigg],
\end{eqnarray}
with $\hat{C} = \hat{B}^{T}$ and finally

\begin{equation}
\hat{D} = R_{0}|00_{c}\rangle\langle 00_{c}| + R_{1}|\psi_{c}^{\pm}\rangle\langle\psi_{c}^{\pm}| + R_{2}|11_{c}\rangle\langle 11_{c}|,
\end{equation}
where we have defined

\begin{equation}
R_{n} = \frac{\Delta - i\gamma/2}{d_{n}}.
\end{equation}
Equipped with the above we finally have all the necessary ingredients to apply the effective operator formalism, via Eqs.(\ref{heffdef})-(\ref{leffdef}), and obtain effective operators for this physical setup. The full set of effective operators, along with the effective Hamiltonian, is given as (\ref{l2arb}) - (\ref{effham}) in Appendix B. All effective operators are of a similar form to $\hat{L}_{1}^{\mathrm{eff}}$, the operator describing effective processes corresponding to a coherent excitation, intermediate process via $H_{NH}^{-1}$ and decay via cavity loss, which is given by 

\begin{eqnarray}\label{l1arb}
\fl\hat{L}_{1}^{\mathrm{eff}} = 2g\Omega\sqrt{\kappa} \bigg[\bigg(\frac{\mathrm{cos}(\theta) + f(\theta)\mathrm{sin}(\theta)}{\tilde{d}_{1}}\bigg) |\psi^{+}\rangle\langle 00 |  
+ \bigg(\frac{f(\theta)\mathrm{cos}(\theta) -\mathrm{sin}(\theta)}{\tilde{d}_{1}}\bigg) |\psi^{-}\rangle\langle 00 | \nonumber\\
+ \bigg(\frac{\mathrm{cos}(\theta) -f(\theta)\mathrm{sin}(\theta)}{\tilde{d}_{2}}\bigg) |11\rangle\langle \psi^{-} | 
+ \bigg(\frac{f(\theta)\mathrm{cos}(\theta) + \mathrm{sin}(\theta)}{\tilde{d}_{2}}\bigg) |11\rangle\langle \psi^{+} | \bigg].
\end{eqnarray}

\section{Parameter Engineering and Restrictions}\label{parameter}

Given expressions (\ref{l1arb}) and (\ref{l2arb})-(\ref{effham}) for the effective operators, it is possible to gain useful insight into the physical processes present in this setup. We wish to prepare the states $|\psi^{+}\rangle$ and $|\psi^{-}\rangle$ for arbitrary values of $\theta$ and it is helpful to begin by identifying all possible effective  pathways in and out of these target states. In this regard, the first thing that is clear from the effective operators is that, if we consider $\hat{H}^{\mathrm{eff}} = \hat{L}_{6}^{\mathrm{eff}} $, then for all $j \in \{1,2,3,4,5,6\}$

\begin{eqnarray}
\mathrm{if} \quad |\nu\rangle = |11\rangle \qquad\qquad\Rightarrow\quad \langle \psi^{\pm} | \hat{L}_{j}^{\mathrm{eff}} | \nu \rangle = 0, \label{firstex}\\
\mathrm{if} \quad |\nu\rangle \in \{|\psi^{\pm}\rangle,|00\rangle\} \quad\nRightarrow\quad\langle \psi^{\pm} | \hat{L}_{j}^{\mathrm{eff}} | \nu \rangle = 0. \label{secondex}
\end{eqnarray}
Expression (\ref{firstex}) shows that it is impossible to move into either target state from the initial state  $|11\rangle$, while expression (\ref{secondex}) shows that movement into either target state from initial states in the set $ \{|\psi^{\pm}\rangle,|00\rangle\}$ is not precluded. In practical quantum optical situations it is not difficult to prepare the state $|00\rangle$ with excellent fidelities, by processes such as Raman cooling \cite{qedcaltech}, and therefore we will assume the initial state of the system as $|00\rangle$. Under this assumption it useful to proceed by examining the differences between effective processes into target states, as opposed to effective processes out of target states, which we would like to suppress. From the effective operators it can be seen that we can write,

\begin{eqnarray}
\langle \psi^{\pm} | \hat{L}_{j}^{\mathrm{eff}} | 00 \rangle = \frac{p^{\pm}_{j}(\tilde{\tau})\xi^{\pm}_{j}(\theta)}{\tilde{d}_{1}},     \label{defn1} \\
\langle \omega | \hat{L}_{j}^{\mathrm{eff}} | \psi^{\pm} \rangle = \frac{\varphi^{(\pm,\omega)}_{j}(\tau)}{\tilde{d}_{2}} ,\label{defn2}
\end{eqnarray}
where $j \in \{1,2,3,4,5,6\}$ indexes the effective operators, the $\pm$ indicates either $+$ or $-$, the final states are from the set $\langle \omega | \in \{\langle\psi^{\pm}|,\langle 11|\}$ and $\tau = \{ g,\delta,\Delta,\kappa,\gamma,\theta,\Omega\}$ and $\tilde{\tau} = \{ g,\delta,\Delta,\kappa,\gamma,\Omega \}$ are parameter sets upon which the functions

\begin{equation} \label{functionset}
\{p^{\pm}_{j}(\tilde{\tau}),\xi^{\pm}_{j}(\theta),\varphi^{(\pm,\omega)}_{j}(\tau)  \}
\end{equation}
are dependent. The functions within the set (\ref{functionset}) are defined by the relationships in (\ref{defn1}) and (\ref{defn2}) and it is crucial for the analysis which follows to note that each function within the set (\ref{functionset}) relates a specific initial state to a specific final state, via a specific effective operator, where the specific initial state, final state and effective operator are encoded in the function name. The full form of the functions within the set (\ref{functionset}) can be shown from the definitions (\ref{defn1}) and (\ref{defn2}), however their full form is complex and not of interest to us at this point. On the contrary, at this stage it is crucial to note, from (\ref{defn1}), that all effective processes into the target state, from the assumed initial state $|00\rangle$,  involve an intermediate process whose strength is proportional to $d_{1}^{-1}$, while (\ref{defn2}) shows that all effective processes out of the target state, into a state $\langle \omega | \in \{\langle\psi^{\pm}|,\langle 11|\}$, involve an intermediate process whose strength is proportional to $d_{2}^{-1}$. Therefore, if it is true that

\begin{equation}\label{arbrestriction}
|p^{\pm}_{j}(\tilde{\tau})\xi^{\pm}_{j}(\theta)| \approx |\varphi^{(\pm,\omega)}_{j}(\tau)|,
\end{equation}
then engineering the strengths of the intermediate propagators such that

\begin{equation}\label{dresarb}
d_{1}^{-1} \gg d_{2}^{-1}  \iff   \tilde{d}_{1}  \ll \tilde{d}_{2},
\end{equation}
will ensure that for all final states $\langle \omega | \in \{\langle\psi^{\pm}|,\langle 11|\}$ and for all effective processes, indexed by $j \in \{1,2,3,4,5,6\}$, it is true that

\begin{equation}\label{requirement}
|\langle \psi^{\pm} | \hat{L}_{j}^{\mathrm{eff}} | 00 \rangle| \gg |\langle \omega | \hat{L}_{j}^{\mathrm{eff}} | \psi^{\pm} \rangle|   .
\end{equation}
Expression (\ref{requirement}) shows that if (\ref{arbrestriction}) and (\ref{dresarb}) are satisfied, then movement into the target state, from assumed initial state $|00\rangle$, is greatly enhanced, while effective movement out of the target state, into any state $\langle \omega | \in \{\langle\psi^{\pm}|,\langle 11|\}$,  is effectively suppressed, regardless of whether the target state is $|\psi^{+}\rangle$ or $|\psi^{-}\rangle$. Mechanisms and requirements for engineering propagator strengths such that (\ref{arbrestriction}) and (\ref{dresarb}) are satisfied will be discussed in detail shortly, however in principal this is achieved through adjustment of system parameters, present in the expressions for effective operators, corresponding to laser driving strengths and detunings.

\begin{figure} 
\begin{center}
\includegraphics[width=0.7\linewidth]{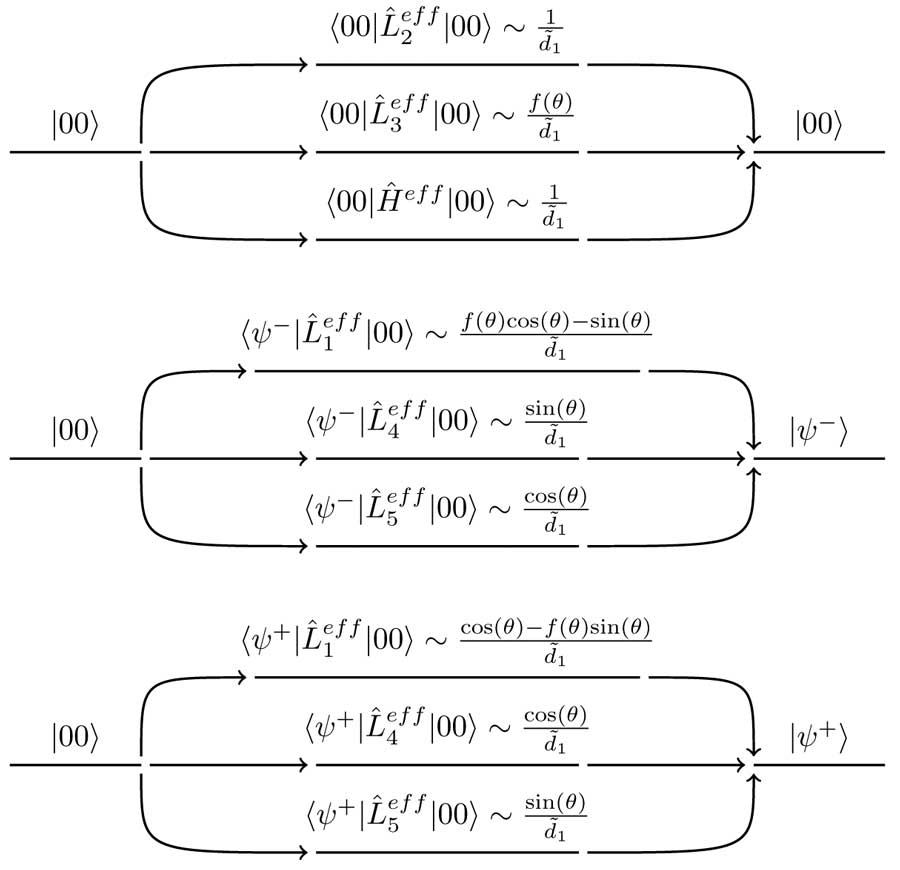} 
\caption{Summary of all effective processes from state $|00\rangle$. It is clear that there are three possible effective processes, each with three possible pathways, where each pathway is facilitated by a different decay mechanism. It is also clear that all effective processes from state $|00\rangle$ are proportional to $d_{1}^{-1}$, and hence involve intermediate processes of the same strength.}\label{arbeff} 
\end{center}
\end{figure}

At this stage it is natural to proceed by considering a mechanism for the preparation of a particular target state, either $|\psi^{+}\rangle$ or $|\psi^{-}\rangle$, assuming the initial state of the system as $|00\rangle$. Figure \ref{arbeff} summarises all effective processes out of the state $|00\rangle$ and details the trigonometric dependence of functions within the set (\ref{functionset}), which are now of importance to our analysis. It is our goal to identify forms for $f(\theta)$, from which laser driving parameters can be chosen such that the desired target state is prepared effectively and reliably.   From Figure \ref{arbeff} and the effective operators (\ref{l1arb}) and (\ref{l2arb})-(\ref{effham}), one finds that if

\begin{equation}\label{choice1}
f(\theta) = -\mathrm{cot}(\theta),
\end{equation}
then

\begin{eqnarray}
\xi^{-}_{1}(\theta) = -\frac{1}{\mathrm{sin}(\theta)}, \label{proa} \\
\xi^{+}_{1}(\theta) = 0. \label{prob}
\end{eqnarray}
From the definition of $\xi^{+}_{1}(\theta)$ and $\xi^{-}_{1}(\theta)$, given in (\ref{defn1}), it can be seen how (\ref{proa}) and (\ref{prob}) show that effective processes, described by $L^{\mathrm{eff}}_{1}$, involving dissipation via cavity loss, will greatly favour movement from $|00\rangle$ into $|\psi^{-}\rangle$, as opposed to movement into $|\psi^{+}\rangle$, for all values of $\theta$. It is also true that

\begin{equation}\label{fornow}
|\xi^{-}_{1}(\theta)|  > |\xi^{-}_{j}(\theta)|   \quad \forall \quad j \in \{4,5\} .
\end{equation} 
Comparison with (\ref{defn1}) shows how (\ref{fornow}) illustrates that, for the function choice given in (\ref{choice1}), cavity loss is the dominant decay mechanism for the production of $|\psi^{-}\rangle$. However, while it is true that for all values of $\theta$,

\begin{equation}
|\xi^{-}_{1}(\theta)| > |\xi^{+}_{4}(\theta)|,
\end{equation}
it is unavoidable that for some values of $\theta$ we have that

\begin{equation}\label{minusthetarestriction}
|\xi^{-}_{1}(\theta)| \approx |\xi^{+}_{5}(\theta)|,
\end{equation}
which is an initial restriction on the values of $\theta$ for which the protocol will work, as in order to completely favour production of $|\psi^{-}\rangle$ as opposed to $|\psi^{+}\rangle$, for a specific value of $\theta$, it is required that

\begin{equation}
|\xi^{-}_{1}(\theta)| > |\xi^{+}_{j}(\theta)|    \quad \forall \quad j \in \{1,4,5\}.
\end{equation}

Similarly, if we wish to prepare the state $|\psi^{+}\rangle$ then one can see, from Figure 3 and the form of the effective operators, that if

\begin{equation}\label{choice2}
f(\theta) = \mathrm{tan}(\theta),
\end{equation}
then

\begin{eqnarray}
\xi^{+}_{1}(\theta) = \frac{1}{\mathrm{cos}(\theta)}, \label{prob1}  \\
\xi^{-}_{1}(\theta) = 0, \label{prob2}
\end{eqnarray}
and 

\begin{equation}
|\xi^{+}_{1}(\theta)|  > |\xi^{+}_{j}(\theta)|   \quad \forall \quad j \in \{4,5\}. 
\end{equation} 
Once again, comparison with the structure of effective processes detailed in (\ref{defn1}), illuminates how (\ref{prob1}) and (\ref{prob2}) show that for laser driving modulation given by (\ref{choice2}), effective processes, described by $L^{\mathrm{eff}}_{1}$, involving dissipation via cavity loss, will be the dominant mechanism for the production of $|\psi^{+}\rangle$, and will greatly favour movement from $|00\rangle$ into $|\psi^{+}\rangle$, as opposed to movement into $|\psi^{-}\rangle$, for all values of $\theta$. However, again it is required that for a specific value of $\theta$, in order to completely favour production  of $|\psi^{+}\rangle$ as opposed to $|\psi^{-}\rangle$, with respect to all decay mechanisms, it is necessary that

\begin{figure} 
\begin{center}
\includegraphics[width=0.7\linewidth]{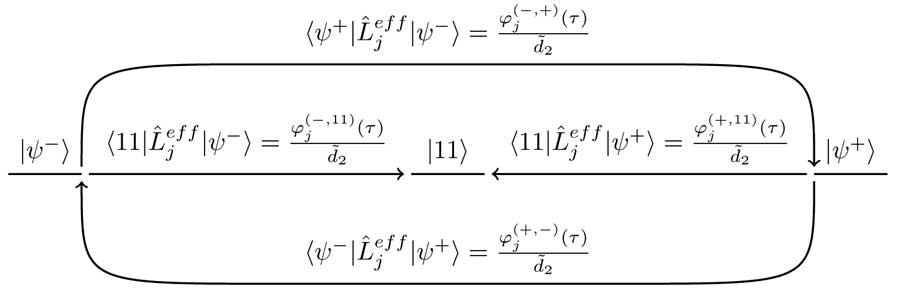} 
\caption{Summary of all effective processes \emph{out} of target states $|\psi^{+}\rangle$ and $|\psi^{-}\rangle$. For each effective process there are multiple pathways, corresponding to different values of  $j$, where each pathway is facilitated by a different decay mechanism. It is also clear that all effective processes out of target states are proportional to $d_{2}^{-1}$, and hence involve intermediate processes of the same strength. It is essential to note which functions relate which final and initial states, via which propagator.}\label{arbeffout} 
\end{center}
\end{figure}

\begin{equation}
|\xi^{+}_{1}(\theta)| > |\xi^{-}_{j}(\theta)|    \quad \forall \quad j \in \{1,4,5\}.
\end{equation}
In this case it is true that

\begin{equation}
|\xi^{+}_{1}(\theta)| > |\xi^{-}_{4}(\theta)|,
\end{equation}
for all values of $\theta$, however it is unavoidable that for some values of $\theta$ we have that

\begin{equation}\label{plusthetarestriction}
|\xi^{-}_{1}(\theta)| \approx |\xi^{+}_{5}(\theta)|,
\end{equation}
a restriction on the values of $\theta$ for which the protocol will work for the production of $|\psi^{+}\rangle$. At this point it is necessary to conduct an in-depth analysis into the extent to which it is possible to achieve the necessary restriction (\ref{arbrestriction}), with the driving modulation function $f(\theta)$ as per (\ref{choice1}) or (\ref{choice2}), depending on the desired target state. From Figure \ref{arbeffout} it is clear that, for the case in which the target state is $|\psi^{-}\rangle$, as decay via cavity loss is the primary mechanism of production, we only require that for each specific value of $\theta$

\begin{equation}\label{minusrequirement}
|p^{-}_{1}(\tilde{\tau})\xi^{-}_{1}(\theta)| \geq |\varphi^{(-,\omega)}_{j}(\tau)|  \quad \forall \quad |\omega\rangle \in \{|\psi^{+}\rangle,|11\rangle\}.
\end{equation}
Similarly, for the case when the target state is $|\psi^{+}\rangle$, we require that for each specific value of $\theta$

\begin{equation}\label{plusrequirement}
|p^{+}_{1}(\tilde{\tau})\xi^{+}_{1}(\theta)| \geq |\varphi^{(+,\omega)}_{j}(\tau)|  \quad \forall \quad |\omega\rangle \in \{|\psi^{-}\rangle,|11\rangle\}.
\end{equation}
As cavity parameters are fixed by co-operativity requirements and the restriction (\ref{dresarb}), the extent to which (\ref{minusrequirement}) and (\ref{plusrequirement}) can be fulfilled is purely dependent on the $\theta$ dependence of $|\varphi^{(-,\omega)}_{j}(\tau)|$ and $|\varphi^{(+,\omega)}_{j}(\tau)|$. For the case when $f(\theta) = -\mathrm{cot}(\theta)$, with $|\psi^{-}\rangle$ as the target state, the following trigonometric terms determine the $\theta$ dependence of $|\varphi^{(-,\omega)}_{j}(\tau)|$ for all relevant $j$ and $|\omega\rangle \in \{|\psi^{+}\rangle,|11\rangle\}$

\begin{eqnarray}\label{minustrigfns1}
f(\theta)\mathrm{cos}(\theta)   \longrightarrow  -\mathrm{cos}(\theta)\mathrm{cot}(\theta),  \\
f(\theta)\mathrm{sin}(\theta)   \longrightarrow  -\mathrm{cos}(\theta), 
\end{eqnarray}
while for the case when $f(\theta) = \mathrm{tan}(\theta)$, with $|\psi^{+}\rangle$ as the target state, the following trigonometric terms determine the $\theta$ dependence of $|\varphi^{(+,\omega)}_{j}(\tau)|$ for all relevant $j$ and $|\omega\rangle \in \{|\psi^{-}\rangle,|11\rangle\}$

\begin{eqnarray}\label{plustrigfns}
f(\theta)\mathrm{cos}(\theta)   \longrightarrow  \mathrm{sin}(\theta), \\
f(\theta)\mathrm{sin}(\theta)   \longrightarrow  \mathrm{sin}(\theta)\mathrm{tan}(\theta).
\end{eqnarray}
Therefore it is clear that, for the case of $f(\theta) = -\mathrm{cot}(\theta)$, with $|\psi^{-}\rangle$ as the target state, the requirement (\ref{minusrequirement}) is strongly fulfilled, i.e.,

\begin{equation}\label{pred1}
|p^{-}_{1}(\tilde{\tau})\xi^{-}_{1}(\theta)| > |\varphi^{(-,\omega)}_{j}(\tau)|  \quad \forall \quad |\omega\rangle \in \{|\psi^{+}\rangle,|11\rangle\},
\end{equation}
when

\begin{equation}
\bigg|\frac{1}{\mathrm{sin}(\theta)}\bigg| > |\mathrm{cos}(\theta)\mathrm{cot}(\theta)| \quad \mathrm{and}  \quad \bigg|\frac{1}{\mathrm{sin}(\theta)}\bigg| > |\mathrm{cos}(\theta)|,
\end{equation}
and weakly fulfilled, i.e.,

\begin{equation}
|p^{-}_{1}(\tilde{\tau})\xi^{-}_{1}(\theta)| \approx |\varphi^{(-,\omega)}_{j}(\tau)|  \quad \forall \quad |\omega\rangle \in \{|\psi^{+}\rangle,|11\rangle\},
\end{equation}
when

\begin{equation}
\bigg|\frac{1}{\mathrm{sin}(\theta)}\bigg| \approx |\mathrm{cos}(\theta)\mathrm{cot}(\theta)| \quad \mathrm{and}  \quad \bigg|\frac{1}{\mathrm{sin}(\theta)}\bigg| \approx |\mathrm{cos}(\theta)|.
\end{equation}
Therefore it is clear that the success of the protocol for the preparation of $|\psi^{-}\rangle$ will vary with $\theta$, in accordance with extent to which (\ref{minusrequirement}) is satisfied. Similarly, for the case of $f(\theta) = \mathrm{tan}(\theta)$, with $|\psi^{+}\rangle$ as the target state, we expect the protocol to work excellently when

\begin{equation}
\bigg|\frac{1}{\mathrm{cos}(\theta)}\bigg| > |\mathrm{sin}(\theta)\mathrm{tan}(\theta)| \quad \mathrm{and}  \quad \bigg|\frac{1}{\mathrm{cos}(\theta)}\bigg| > |\mathrm{sin}(\theta)|,
\end{equation}
and less successfully when

\begin{equation}\label{pred2}
\bigg|\frac{1}{\mathrm{cos}(\theta)}\bigg| \approx |\mathrm{sin}(\theta)\mathrm{tan}(\theta)| \quad \mathrm{and}  \quad \bigg|\frac{1}{\mathrm{cos}(\theta)}\bigg| \approx |\mathrm{sin}(\theta)|.
\end{equation}
These characteristics, for both target states, are clearly seen in Figure \ref{results3} in Section \ref{ressec}. Finally, it is necessary to demonstrate a method for choosing cavity parameters such that the requirement (\ref{dresarb}) is fulfilled. In order to do this we introduce the notation

\begin{equation}
g = y,  \qquad\quad \delta = \tilde{\delta}y, \qquad  \Delta = \tilde{\Delta}y,
\end{equation}
\begin{equation}
\Omega = \tilde{\Omega} x, \qquad\kappa = \tilde{\kappa}x, \qquad \gamma = \tilde{\gamma}x,
\end{equation}
where $y = \alpha x$, $\alpha \approx 10$ and $(\tilde{\delta},\tilde{\Delta},\tilde{\Omega},\tilde{\kappa}, \tilde{\gamma}) = \mathcal{O}(1)$ enforce the correct scale of each parameter. Utilizing this notation one finds

\begin{equation}
\tilde{d}_{j} =x^2 \Big[4\alpha^2 \big( j - \tilde{\delta}\tilde{\Delta}\big) + 2i\alpha \big(\tilde{\delta}\tilde{\gamma} + \tilde{\Delta}\tilde{\kappa} \big)  + \tilde{\gamma}\tilde{\kappa} \Big],
\end{equation}
and hence to leading order

\begin{equation}\label{leadingorder}
\tilde{d}_{j} \approx 4 x^2 \alpha^2 \big( j - \tilde{\delta}\tilde{\Delta}  \big),
\end{equation}
such that that a parameter choice $\tilde{\delta}\tilde{\Delta} = 1$, practically achievable via the use of specific laser detunings, yields

\begin{eqnarray}
\tilde{d}_{1} \approx 0, \\
\tilde{d}_{2}  \approx 4 x^2 \alpha^2,
\end{eqnarray}
effectively yielding $|\tilde{d}_1|^2 \ll |\tilde{d}_2|^2$ as required.

\section{Numerical Analysis and results}\label{ressec}

The effective master equation (\ref{effmastereq}) corresponding to effective operators (\ref{l1arb}),(\ref{l2arb})-(\ref{l5arb}) is extremely non-trivial to solve exactly, and hence for this protocol we utilise numerical analysis to analyse the characteristic behaviour of the scheme. We are primarily interested in the behaviour of the protocol, for both possible target states, with respect to $\theta$ and $C$, where 

\begin{equation}
C = \frac{g^{2}}{\kappa\gamma},
\end{equation}
is the cavity cooperativity, an invariant measure of the quality of a cavity QED setup \cite{qedreview3}.

\begin{figure} 
\begin{center}
\includegraphics[width=0.9\linewidth]{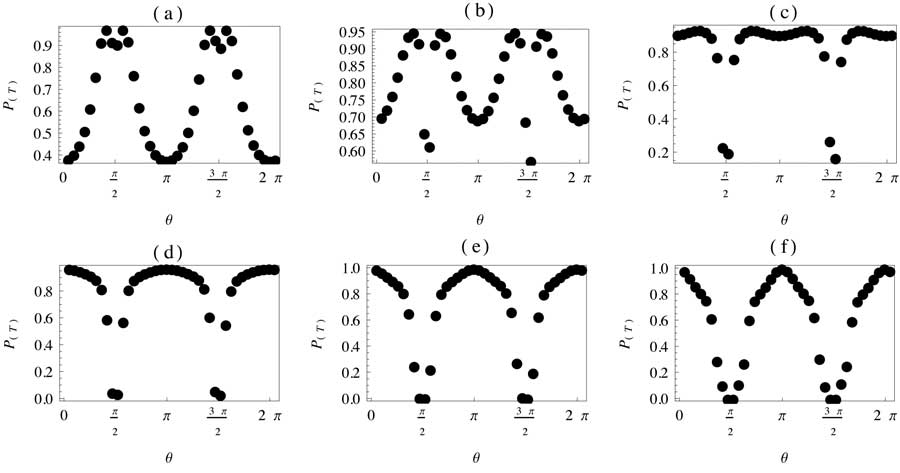} 
\caption{Population of $|\psi^{+}\rangle$ at varying time $T$, represented by $P_{(T)}$, for all $\theta \in [0,2\pi]$, with $f(\theta) = \mathrm{tan}(\theta)$. All plots are in the regime of weak driving with $\tilde{\Omega} = 1/10$, and detunings $(\tilde{\delta},\tilde{\Delta}) = (1/2,2)$ satisfying restriction (\ref{dresarb}). Furthermore, all plots are at $C=200$ with $(\tilde{\kappa},\tilde{\gamma}) = (1,1/2)$. All times are in relevant units of $1/g$, where for (a) $T=2000$ (b) $T=5000$ (c) $T=10 000$ (d) $T=15000$ (e) $T=25000$ (f) $T=50000$. } \label{thetadecays} 
\end{center}
\end{figure}

\begin{figure} 
\begin{center}
\includegraphics[width=0.9\linewidth]{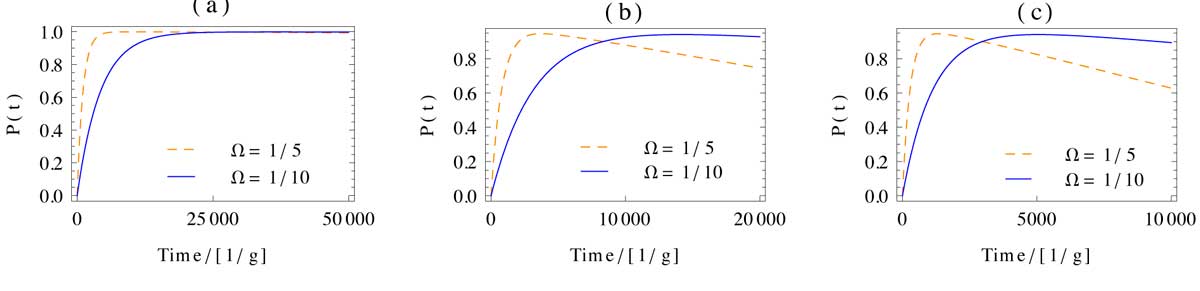} 
\caption{Evolutions of target state $|\psi^{+}\rangle$ with time, with $f(\theta) = \mathrm{tan}(\theta)$ and for different specific values of $\theta$ and different coherent driving strengths. All plots are in the regime of weak driving with $\tilde{\Omega} = 1/10$ or $\tilde{\Omega} = 1/5$, and detunings $(\tilde{\delta},\tilde{\Delta}) = (1/2,2)$ satisfying restriction (\ref{dresarb}). Furthermore, all plots are at $C=200$ with $(\tilde{\kappa},\tilde{\gamma}) = (1,1/2)$. The values of $\theta$ for each plot are (a) $\theta = \theta_{+}  = \pi$ (b) $\theta =  \theta_{+} \pm \pi/6$ (c) $\theta =  \theta_{+}  \pm \pi/3$. Plots of the target state $|\psi^{-}\rangle$ with time, with $f(\theta) = -\mathrm{cot}(\theta)$ and all other parameters as above, are identical to figures (a)-(c) if $\theta_{+}$ is replaced with $\theta_{-}$.}\label{evolutions} 
\end{center}
\end{figure}

Figures \ref{thetadecays} and \ref{evolutions} allow us insight into the evolutions of the target states, for different values of $\theta$, at set values of cooperativity. From Figure \ref{thetadecays}, which displays the evolution of the target state $|\psi^{+}\rangle$ when $f(\theta) = \mathrm{tan}(\theta)$, it is clear that the behaviour of the protocol is periodic with respect to $\theta$, with a period of $\pi$. Furthermore, the behaviour of the protocol, again with respect to $\theta$, is symmetric about the midpoint of any period. If we define $\theta^{\pm}_{m}$ as the value of $\theta$ for which the population of $|\psi^{\pm}\rangle$ is a maximum over $\theta$ as $T \rightarrow \infty$, then it is clear that for the protocol with $|\psi^{+}\rangle$ as a target state we have that

\begin{figure} 
\begin{center}
\includegraphics[width=0.9\linewidth]{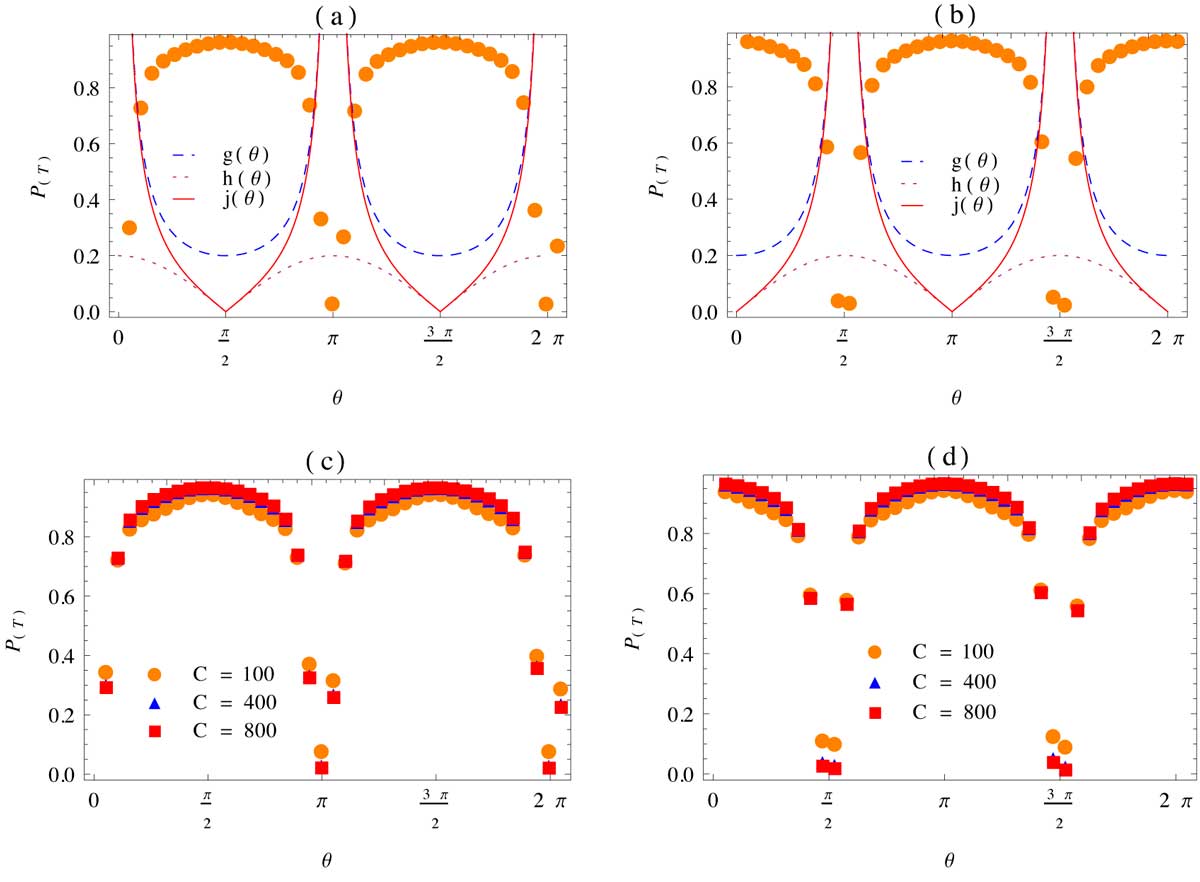} 
\caption{Population at time $T = 15000$, in units of $1/g$, for (b,d) $|\psi^{+}\rangle$ with $f(\theta) = \mathrm{tan}(\theta)$ (a,c) $|\psi^{-}\rangle$ with $f(\theta) = -\mathrm{cot}(\theta)$ and for all $\theta \in [0,2\pi]$. In (a) and (b) populations are overlaid with scaled trigonometric terms governing $\theta$ dependence of effective operators. The trigonometric functions are as per (\ref{plustrigfns}) and (\ref{minustrigfns1}), scaled appropriately with (a) $g(\theta) = |\mathrm{cosec}(\theta)|/5$, $h(\theta) = |\mathrm{cos}(\theta)|/5$, $j(\theta) = |\mathrm{cot}(\theta)|/5$ (b) $g(\theta) = |\mathrm{sec}(\theta)|/5$, $h(\theta) = |\mathrm{sin}(\theta)|/5$, $j(\theta) = |\mathrm{tan}(\theta)|/5$.  In (c) and (d) populations are shown at varying cooperativities, where in (c) and (d) $\tilde{\gamma} = 1/2$ with cooperativity varied through $\tilde{\kappa}$. In (a) and (b) cooperativity is set at $C=200$ with $(\tilde{\kappa},\tilde{\gamma}) = (1,1/2)$.  All plots are in the regime of weak driving with $\tilde{\Omega} = 1/10$, and detunings $(\tilde{\delta},\tilde{\Delta}) = (1/2,2)$ satisfying restriction (\ref{dresarb}). }\label{results3} 
\end{center}
\end{figure}

\begin{equation}
\theta^{+}_{m} = \pi.
\end{equation}
One can also see from Figure \ref{thetadecays} that if we consider a period with $\theta^{+}_{m}$ as a midpoint, then the behaviour of the protocol is best, in the sense that the target state is stable for extremely long times, for $\theta =\theta^{+}_{m} $, while the effectiveness of the protocol, with respect to stability of the target state, decreases symmetrically outwards towards $\theta =\theta^{+}_{m} \pm \pi/2$. This behaviour is in direct accordance with the analytical restrictions of (\ref{pred1}) - (\ref{pred2}), as can easily be seen from Figure \ref{results3}. One can also see from Figure \ref{results3} that the protocol for the preparation of $|\psi^{-}\rangle$, with $f(\theta) = -\mathrm{cot}(\theta)$, behaves identically, with respect to periodicity and symmetry, to the protocol for the preparation of  $|\psi^{+}\rangle$, except with

\begin{equation}
\theta^{-}_{m} = \frac{\pi}{2}.
\end{equation}
which is again perfectly aligned with the theoretical analysis of (\ref{pred1}) - (\ref{pred2}). For comparison with previously suggested schemes \cite{DPREPQED1, DPREPQED2}, Figure \ref{evolutions} allows a more detailed insight into the evolutions of the target states and illustrates that for $\theta =\theta^{+}_{m}$, with $f(\theta) = -\mathrm{cot}(\theta)$, the protocol replicates the results of \cite{DPREPQED1,DPREPQED2} in the sense that the maximally entangled Bell State of two-qubits is prepared as the steady state of the system, with fidelities of near unity, and with cooperativities within reach of present technologies \cite{qedcaltech,qedreview3}. Moreover, both $|\psi^{+}\rangle$ and $|\psi^{-}\rangle$ can be prepared, with fidelities of near unity and achievable cooperativities, stable for times $T = \mathcal{O}(10 000)$, for values of $\theta$ within $\theta^{\pm}_{m} \pm  \pi/3$, allowing for the preparation of the large majority of entangled states of two-qubits. 

It is important to note that we are restricting ourselves to the regime of weak driving, in which adiabatic elimination is applicable and the accuracy of the effective operator formalism has been well established \cite{effop, DPREPQED1}. Therefore, as it is necessary to utilise individual addressing of atoms, with the coherent driving strength of the second atom equal to $f(\theta)\Omega$, we are restricted to values of $\theta$ for which $|f(\theta)| = \mathcal{O}(1)$. Therefore, despite the characteristics of the protocol displayed in the previous figures, the asymptotic behaviour of the necessary choices for $f(\theta)$ would still imply a restriction to $\theta^{\pm}_{m} \pm  \pi/3$, for the preparation of both $|\psi^{+}\rangle$ and $|\psi^{-}\rangle$. As per previously suggested schemes, as can be seen in Figure \ref{evolutions}, increased driving allows one to obtain maximum fidelities more rapidly, but at the expense of target state stability. 

Finally, Figure \ref{results3} shows that for the protocol suggested here the scaling of the protocol behaviour, with respect to cooperativity, is excellent. As can be seen from Figure \ref{results3}, excellent fidelities can be obtained with $C=100$, corresponding to currently available cavities \cite{qedreview3}, while at increased cooperativities, such as $C = 800$, even better fidelities are obtainable.

\section{Conclusions and outlook}

We have presented a scheme for the dissipative preparation of generalised Bell states, defined by (\ref{arbtarg1})-(\ref{arbtarg2}), effectively allowing for the preparation of arbitrary two-qubit states and therefore excellent control over a two-qubit cavity QED system. In the presented schemes the dissipative processes of spontaneous emission and cavity loss are no longer undesirable, but essential to the required dynamics. Furthermore, contrary to a typically active approach to state preparation, our approach allows for the system to be prepared, such that when left alone the system ``cools" into the desired target state, with dissipation as the driving mechanism.

The proposed scheme achieves extremely long lived target states, of excellent fidelities, utilising cavity cooperativities corresponding to currently available optical cavities \cite{qedreview3}. In addition, the protocol exhibits excellent scaling characteristics, with respect to cooperativity, such that even better results may be achieved as current experimental capabilities develop.

For practical purposes it may be necessary to achieve a \emph{true} steady state. While we are interested in this paper in truly dissipative dynamics, in which the systems cools into the desired state, a more active approach, involving some measure of feedback or measurement as utilised in \cite{DPREPREAL2}, may be succesfully adopted, as per currently ongoing research. Furthermore, in a realistic experiment one would like to characterise the states which are being produced. For general systems, to which the methods and mathematical formalism of this work may be applicable, the witness methods of \cite{witness} offer a possible means for state characterisation, while for QED setups, as discussed in this work, atomic state tomography \cite{tom} may be utilised for the characterisation and verification of these results.

\ack This work is based upon research supported by the
South African Research Chair Initiative of the Department of Science and Technology and National Research
Foundation. R. S. also acknowledges support from the National Institute for Theoretical Physics.

\appendix
\section{Hamiltonian and Lindblad operators in basis $B$}

After a transformation into the basis $B$, as per (\ref{basis}), the Hamiltonian for the excited state manifold is given by,

\begin{equation}\label{transH}
\fl \hat{H}_{\mathrm{e}} = \Delta \Big(\sum_{i=1}^{2} |\psi_{i}^{\pm}\rangle\langle\psi_{i}^{\pm}| \Big) + \delta\Big( |00_{c}\rangle\langle 00_{c}| + |11_{c}\rangle\langle 11_{c}| + |\psi_{c}^{\pm}\rangle\langle\psi_{c}^{\pm}|\Big) + \hat{H}_{\mathrm{ac}}, 
\end{equation}
with the atom-cavity interaction term

\begin{equation}
\fl \hat{H}_{\mathrm{ac}} = g\Big(|\psi_{c}^{\pm}\rangle\langle\psi_{0}^{\pm}| + |11_{c}\rangle\Big[ \langle \psi_{1}^{+}|\big(\mathrm{cos}(\theta) + \mathrm{sin}(\theta)\big) +  \langle \psi_{1}^{-}|\big(\mathrm{cos}(\theta) - \mathrm{sin}(\theta)\big)\Big]     \Big) + \mathrm{H.c},
\end{equation}
where we have utilized an implied summation notation,

\begin{equation}
|\psi_{j}^{\pm}\rangle\langle\psi_{j}^{\pm}| = |\psi_{j}^{+}\rangle\langle\psi_{j}^{+}| + |\psi_{j}^{-}\rangle\langle\psi_{j}^{-}|.
\end{equation}
The perturbative excitation term of the Hamiltonian takes the form

\begin{eqnarray}
\fl \hat{W}_{+} = \frac{\Omega}{2}\Big[   \big(\mathrm{cos}(\theta) + f(\theta)\mathrm{sin}(\theta)\big)|\psi_{0}^{+}\rangle\langle 00| 
 +  \big(f(\theta)\mathrm{cos}(\theta) - \mathrm{sin}(\theta)\big)|\psi_{0}^{-}\rangle \langle 00|  
+(\big(\mathrm{cos}^2(\theta)  \nonumber\\- f(\theta)\mathrm{sin}^2(\theta)\big)|\psi_{1}^{+}\rangle \langle \psi^{-}|  
 + \big(\mathrm{cos}(\theta)\mathrm{sin}(\theta) + f(\theta)\mathrm{cos}(\theta)\mathrm{sin}(\theta)\big)|\psi_{1}^{+}\rangle \langle \psi^{+}| 
 \nonumber\\-\big(\mathrm{cos}(\theta)\mathrm{sin}(\theta) + f(\theta)\mathrm{cos}(\theta)\mathrm{sin}(\theta)\big)|\psi_{1}^{-}\rangle \langle \psi^{+}|  
\nonumber\\+ \big(f(\theta)\mathrm{cos}^2(\theta) - \mathrm{sin}^2(\theta)\big)|\psi_{1}^{-}\rangle \langle \psi^{-}| \Big], 
\end{eqnarray}
while the perturbative de-excitation is given by $\hat{W}_{-} = \hat{W}_{+}^{\dagger} $ and $\hat{H}_{g} = 0$.  After the same basis transformation the Lindblad operator describing cavity loss becomes

\begin{equation}
\hat{L}_{1} = \hat{L}_{\kappa} = \sqrt{\kappa}\Big( |00\rangle\langle 00_{c}| + |11\rangle\langle 11_{c}| + |\psi^{\pm}\rangle\langle\psi_{c}^{\pm}|\Big),
\end{equation}
while the Lindblad operators describing spontaneous emission of both atoms into $|0\rangle$ become

\begin{eqnarray}
\fl\hat{L}_{2} = \hat{L}_{[\gamma,(1,0)]} = |00\rangle\langle \psi_{0}^{+}| \bigg(\sqrt{\frac{\gamma}{2}}\mathrm{cos}(\theta) \bigg)  
 -   |00\rangle\langle \psi_{0}^{-}| \bigg(\sqrt{\frac{\gamma}{2}}\mathrm{sin}(\theta) \bigg) \nonumber\\
 + |\psi^{-}\rangle\langle \psi_{1}^{+}| \bigg(\sqrt{\frac{\gamma}{2}}\mathrm{cos}^2(\theta) \bigg)
 + |\psi^{+}\rangle\langle \psi_{1}^{+}| \bigg(\sqrt{\frac{\gamma}{2}}\mathrm{cos}(\theta)\mathrm{sin}(\theta) \bigg) \nonumber\\
- |\psi^{-}\rangle\langle \psi_{1}^{-}| \bigg(\sqrt{\frac{\gamma}{2}}\mathrm{cos}(\theta)\mathrm{sin}(\theta) \bigg)
- |\psi^{+}\rangle\langle \psi_{1}^{-}| \bigg(\sqrt{\frac{\gamma}{2}}\mathrm{sin}^2(\theta) \bigg), 
\end{eqnarray}

\begin{eqnarray}
\fl \hat{L}_{3} = \hat{L}_{[\gamma,(2,0)]} = |00\rangle\langle \psi_{0}^{+}| \bigg(\sqrt{\frac{\gamma}{2}}\mathrm{sin}(\theta) \bigg)  
 +   |00\rangle\langle \psi_{0}^{-}| \bigg(\sqrt{\frac{\gamma}{2}}\mathrm{cos}(\theta) \bigg) \nonumber\\
- |\psi^{-}\rangle\langle \psi_{1}^{+}| \bigg(\sqrt{\frac{\gamma}{2}}\mathrm{sin}^2(\theta) \bigg)
 + |\psi^{+}\rangle\langle \psi_{1}^{+}| \bigg(\sqrt{\frac{\gamma}{2}}\mathrm{cos}(\theta)\mathrm{sin}(\theta) \bigg) \nonumber\\
 - |\psi^{-}\rangle\langle \psi_{1}^{-}| \bigg(\sqrt{\frac{\gamma}{2}}\mathrm{cos}(\theta)\mathrm{sin}(\theta) \bigg)
+ |\psi^{+}\rangle\langle \psi_{1}^{-}| \bigg(\sqrt{\frac{\gamma}{2}}\mathrm{cos}^2(\theta) \bigg), 
\end{eqnarray}
and the Lindblad operators describing spontaneous emission of both atoms into $|1\rangle$ become

\begin{eqnarray}
\fl\hat{L}_{4} = \hat{L}_{[\gamma,(1,1)]} = |11\rangle\langle \psi_{1}^{+}| \bigg(\sqrt{\frac{\gamma}{2}}\mathrm{cos}(\theta) \bigg)  
 -   |11\rangle\langle \psi_{1}^{-}| \bigg(\sqrt{\frac{\gamma}{2}}\mathrm{sin}(\theta) \bigg) \nonumber\\
 - |\psi^{-}\rangle\langle \psi_{0}^{+}| \bigg(\sqrt{\frac{\gamma}{2}}\mathrm{cos}(\theta)\mathrm{sin}(\theta) \bigg)
 + |\psi^{+}\rangle\langle \psi_{0}^{+}| \bigg(\sqrt{\frac{\gamma}{2}}\mathrm{cos}^2(\theta) \bigg) \nonumber\\
 + |\psi^{-}\rangle\langle \psi_{0}^{-}| \bigg(\sqrt{\frac{\gamma}{2}}\mathrm{sin}^2(\theta) \bigg)
- |\psi^{+}\rangle\langle \psi_{0}^{-}| \bigg(\sqrt{\frac{\gamma}{2}}\mathrm{cos}(\theta)\mathrm{sin}(\theta) \bigg),
\end{eqnarray}

\begin{eqnarray}\label{transL5}
\fl\hat{L}_{5} = \hat{L}_{[\gamma,(2,1)]} = |11\rangle\langle \psi_{1}^{+}| \bigg(\sqrt{\frac{\gamma}{2}}\mathrm{sin}(\theta) \bigg)  
 +   |11\rangle\langle \psi_{1}^{-}| \bigg(\sqrt{\frac{\gamma}{2}}\mathrm{cos}(\theta) \bigg) \nonumber\\
 + |\psi^{-}\rangle\langle \psi_{0}^{+}| \bigg(\sqrt{\frac{\gamma}{2}}\mathrm{cos}(\theta)\mathrm{sin}(\theta) \bigg)
 + |\psi^{+}\rangle\langle \psi_{0}^{+}| \bigg(\sqrt{\frac{\gamma}{2}}\mathrm{sin}^2(\theta) \bigg) \nonumber\\
 + |\psi^{-}\rangle\langle \psi_{0}^{-}| \bigg(\sqrt{\frac{\gamma}{2}}\mathrm{cos}^2(\theta) \bigg)
+ |\psi^{+}\rangle\langle \psi_{0}^{-}| \bigg(\sqrt{\frac{\gamma}{2}}\mathrm{cos}(\theta)\mathrm{sin}(\theta) \bigg). 
\end{eqnarray}
\section{Effective operators and effective Hamiltonian}

The effective operators $\hat{L}_{2}^{\mathrm{eff}}$ and $\hat{L}_{3}^{\mathrm{eff}}$, describing effective processes corresponding to a coherent excitation, intermediate process via $H_{NH}^{-1}$ and decay via spontaneous emission into level $|0\rangle$, are given by

\begin{eqnarray}\label{l2arb}
\fl\hat{L}^{\mathrm{eff}}_{2} = \bigg(\frac{i\sqrt{\gamma}\Omega(2i\delta + \kappa)}{\sqrt{2}\tilde{d}_{1}}\bigg)|00\rangle\langle 00| \nonumber\\
+ \bigg(\frac{i\sqrt{\gamma}\Omega\mathrm{cos}(\theta)\big[\tilde{d}_{1}\mathrm{cos}(\theta) + 4g^2f(\theta)\mathrm{sin}(\theta) \big]}{\sqrt{2}(\gamma + 2i\Delta)\tilde{d}_{2}}\bigg)|\psi^{-}\rangle\langle \psi^{-}| \nonumber\\
+ \bigg(\frac{i\sqrt{\gamma}\Omega\mathrm{sin}(\theta)\big[\tilde{d}_{1}\mathrm{cos}(\theta) + 4g^2f(\theta)\mathrm{sin}(\theta) \big]}{\sqrt{2}(\gamma + 2i\Delta)\tilde{d}_{2}}\bigg)|\psi^{+}\rangle\langle \psi^{-}| \nonumber\\
+ \bigg(\frac{i\sqrt{\gamma}\Omega\mathrm{sin}(\theta)\big[\tilde{d}_{1}\mathrm{sin}(\theta) - 4g^2f(\theta)\mathrm{cos}(\theta) \big]}{\sqrt{2}(\gamma + 2i\Delta)\tilde{d}_{2}}\bigg)|\psi^{+}\rangle\langle \psi^{+}| \nonumber\\
+ \bigg(\frac{i\sqrt{\gamma}\Omega\mathrm{cos}(\theta)\big[\tilde{d}_{1}\mathrm{sin}(\theta) - 4g^2f(\theta)\mathrm{cos}(\theta) \big]}{\sqrt{2}(\gamma + 2i\Delta)\tilde{d}_{2}}\bigg)|\psi^{-}\rangle\langle \psi^{+}|, 
\end{eqnarray}

\begin{eqnarray}\label{l3arb}
\fl\hat{L}^{\mathrm{eff}}_{3} = \bigg(\frac{i\sqrt{\gamma}\Omega(2i\delta + \kappa)f(\theta)}{\sqrt{2}\tilde{d}_{1}}\bigg)|00\rangle\langle 00|  \nonumber\\
+ \bigg(\frac{i\sqrt{\gamma}\Omega\mathrm{sin}(\theta)\big[\tilde{d}_{1}f(\theta)\mathrm{sin}(\theta) + 4g^2\mathrm{cos}(\theta) \big]}{\sqrt{2}(\gamma + 2i\Delta)\tilde{d}_{2}}\bigg)|\psi^{-}\rangle\langle \psi^{-}| \nonumber\\
+ \bigg(\frac{i\sqrt{\gamma}\Omega\mathrm{cos}(\theta)\big[\tilde{d}_{1}f(\theta)\mathrm{sin}(\theta) + 4g^2\mathrm{cos}(\theta) \big]}{\sqrt{2}(\gamma + 2i\Delta)\tilde{d}_{2}}\bigg)|\psi^{+}\rangle\langle \psi^{-}| \nonumber\\
+ \bigg(\frac{i\sqrt{\gamma}\Omega\mathrm{cos}(\theta)\big[\tilde{d}_{1}f(\theta)\mathrm{cos}(\theta) - 4g^2\mathrm{sin}(\theta) \big]}{\sqrt{2}(\gamma + 2i\Delta)\tilde{d}_{2}}\bigg)|\psi^{+}\rangle\langle \psi^{+}| \nonumber\\
+ \bigg(\frac{i\sqrt{\gamma}\Omega\mathrm{sin}(\theta)\big[4g^2\mathrm{sin}(\theta) -\tilde{d}_{1}f(\theta)\mathrm{cos}(\theta)  \big]}{\sqrt{2}(\gamma + 2i\Delta)\tilde{d}_{2}}\bigg)|\psi^{-}\rangle\langle \psi^{+}|. 
\end{eqnarray}
The final effective operators $\hat{L}_{4}^{\mathrm{eff}}$ and $\hat{L}_{5}^{\mathrm{eff}}$, describing effective processes corresponding to a coherent excitation, intermediate process via $H_{NH}^{-1}$ and decay via spontaneous emission into level $|1\rangle$, have the form

\begin{eqnarray}\label{l4arb}
\fl\hat{L}^{\mathrm{eff}}_{4} = \bigg(\frac{i\sqrt{\gamma}\Omega\mathrm{sin}(\theta)(2i\delta +\kappa)}{\sqrt{2}\tilde{d}_{1}}\bigg)|\psi^{-}\rangle\langle 00| \nonumber\\
- \bigg(\frac{i\sqrt{\gamma}\Omega\mathrm{cos}(\theta)(2i\delta +\kappa)}{\sqrt{2}\tilde{d}_{1}}\bigg)|\psi^{+}\rangle\langle 00| \nonumber\\
+ \bigg(\frac{i\sqrt{\gamma}\big[\tilde{d}_{1}\mathrm{cos}(\theta) + 4g^2f(\theta)\mathrm{sin}(\theta) \big]}{\sqrt{2}(\gamma + 2i\Delta)\tilde{d}_{2}}\bigg)|11\rangle\langle \psi^{-}| \nonumber\\
+ \bigg(\frac{i\sqrt{\gamma}\big[\tilde{d}_{1}\mathrm{sin}(\theta) - 4g^2f(\theta)\mathrm{cos}(\theta) \big]}{\sqrt{2}(\gamma + 2i\Delta)\tilde{d}_{2}}\bigg)|11\rangle\langle \psi^{+}|, 
\end{eqnarray}

\begin{eqnarray}\label{l5arb}
\fl\hat{L}^{\mathrm{eff}}_{5} = \bigg(\frac{i\sqrt{\gamma}\Omega f(\theta)\mathrm{cos}(\theta)(2i\delta +\kappa)}{\sqrt{2}\tilde{d}_{1}}\bigg)|\psi^{-}\rangle\langle 00|  \nonumber\\
+ \bigg(\frac{i\sqrt{\gamma}\Omega f(\theta)\mathrm{sin}(\theta)(2i\delta +\kappa)}{\sqrt{2}\tilde{d}_{1}}\bigg)|\psi^{+}\rangle\langle 00| \nonumber\\
- \bigg(\frac{i\sqrt{\gamma}\big[\tilde{d}_{1}f(\theta)\mathrm{sin}(\theta) + 4g^2\mathrm{cos}(\theta) \big]}{\sqrt{2}(\gamma + 2i\Delta)\tilde{d}_{2}}\bigg)|11\rangle\langle \psi^{-}| \nonumber\\
+ \bigg(\frac{i\sqrt{\gamma}\big[\tilde{d}_{1}f(\theta)\mathrm{cos}(\theta) - 4g^2\mathrm{sin}(\theta) \big]}{\sqrt{2}(\gamma + 2i\Delta)\tilde{d}_{2}}\bigg)|11\rangle\langle \psi^{+}|. 
\end{eqnarray}
Finally, the effective Hamiltonian corresponding to unitary processes consisting of a coherent excitation, intermediate propagation via $H_{NH}^{-1}$ and a coherent de-excitation, is found to be

\begin{eqnarray}\label{effham}
\fl\hat{H}^{\mathrm{eff}} = \bigg(\frac{\Omega^{2}h_{(1)}(g,\delta,\Delta,\kappa,\gamma,\theta)}{|d_{1}|^2}\bigg)|00\rangle\langle 00| \nonumber\\
+ \bigg(\frac{\Omega^{2}h_{(2)}(g,\delta,\Delta,\kappa,\gamma,\theta)}{|d_{2}|^2}\bigg)|\psi^{-}\rangle\langle \psi^{-}| \nonumber\\
+ \bigg(\frac{\Omega^{2}h_{(3)}(g,\delta,\Delta,\kappa,\gamma,\theta)}{|d_{2}|^2}\bigg)|\psi^{+}\rangle\langle \psi^{-}| \nonumber\\
+ \bigg(\frac{\Omega^{2}h_{(4)}(g,\delta,\Delta,\kappa,\gamma,\theta)}{|d_{2}|^2}\bigg)|\psi^{+}\rangle\langle \psi^{+}|\nonumber\\
+ \bigg(\frac{\Omega^{2}h_{(3)}(g,\delta,\Delta,\kappa,\gamma,\theta)}{|d_{2}|^2}\bigg)|\psi^{-}\rangle\langle \psi^{+}| ,
\end{eqnarray}
where $h_{(i)}$ are complicated functions whose exact form is explored in Section \ref{parameter}.

\end{document}